\documentclass[a4paper,11pt]{article}
\usepackage{amsfonts}
\usepackage{amsmath,amssymb,amscd}
\usepackage{appendix,marginnote,tikz,pgf,mathtools}
\usepackage{comment}
\usepackage{hyperref}

\usepackage{color}

\usepackage[numbers]{natbib}
\def\ll{\left\langle}
\def\rr{\right\rangle}

\def\bean#1\enan{\begin{align*}#1\end{align*}}
\def\bea#1\ena{\begin{align}#1\end{align}}

\def\bea#1\ena{\begin{align}#1\end{align}}
\def\beaa#1\enaa{\begin{align}\begin{aligned}#1\end{aligned}\end{align}}
\def\bean#1\enan{\begin{align*}#1\end{align*}} 

\def\pd{\partial}

\def\matt[#1,#2,#3,#4]{\left(%
\begin{array}{cc} #1 & #2 \\ #3 & #4 \end{array} \right)}
\def\vect[#1,#2]{\left(%
\begin{array}{cc} #1 \\ #2  \end{array} \right)}
\def\tvect[#1,#2]{\left(%
\begin{array}{cc} #1 & #2  \end{array} \right)}
\def\ket[#1]{\left| #1 \right\rangle}
\def\bra[#1]{\left\langle #1 \right|}
\def\brak[#1,#2]{\left\langle #1 | #2 \right\rangle}
\def\pair[#1,#2]{\left\langle #1 , #2 \right\rangle}

\def\be{\begin{equation}}
\def\ee{\end{equation}}
\def\ben{\begin{equation*}}
\def\een{\end{equation*}}

\def\pd{\partial}
\usepackage{mathtools}

\newcommand{\DeclareAutoPairedDelimiter}[3]{%
	\expandafter\DeclarePairedDelimiter\csname Auto\string#1\endcsname{#2}{#3}%
	\begingroup\edef\x{\endgroup
		\noexpand\DeclareRobustCommand{\noexpand#1}{%
			\expandafter\noexpand\csname Auto\string#1\endcsname*}}%
	\x}

\baselineskip=\normalbaselineskip

\begin{document} 

\begin{center}

{\Large \bf  
 Open KdV hierarchy and minimal gravity on disk
}
\vskip 1cm
{\large Aditya Bawane\footnote{abawane@sogang.ac.kr}, 
Hisayoshi Muraki\footnote{hmuraki@sogang.ac.kr}
 and Chaiho Rim\footnote{rimpine@sogang.ac.kr}
}\\
{Department of Physics, 
Sogang University, Seoul 04107, Korea}
 \end{center}

\vskip 10mm

\begin{abstract}

We show that  the minimal gravity of Lee-Yang series on disk 
 is a solution to the open KdV hierarchy proposed for the 
intersection theory on the moduli space of Riemann surfaces with boundary. 
\end{abstract}

\flushbottom

\section{Introduction} 

The generating function of the intersection theory on the moduli space of Riemann surfaces had been conjectured to satisfy the KdV hierarchy together with the string equation \cite{Witten}. It was shown in \cite{DVV, MMM} that Witten's conjecture is equivalent to the Virasoro constraints. On the other hand, minimal gravity is  2-dimensional quantum gravity  coupled with minimal conformal matter 
so that the resulting theory still remains conformal and topological $(c=0)$ \cite{Kostov, Kazakov, KPZ}, and also obeys the KdV hierarchy \cite{Doug, Ginsparg}. As a result, KdV hierarchy plays an important role 
in understanding intersection theory as well as topological gravity.

Recently, the generating function for intersection theory on the moduli space of Riemann surfaces with boundary has been conjectured to satisfy the so-called open KdV hierarchy which contains the KdV flow parameters as well as new flow parameters related with the boundary \cite{PST}. It has also been checked that the open KdV hierarchy can be
represented in modified form of Virasoro constraints \cite{Buryak}.
It is naturally expected that the open KdV hierarchy also describes minimal gravity on a disk. We check this expectation in this letter.

In section 2, we summarize minimal gravity on sphere and its connection with KdV hierarchy 
and in section 3, open KdV hierarchy is explicitly checked for the free energy of minimal gravity on a disk. 
Section 4 is the conclusion.

\section{Minimal gravity on sphere and KdV hierarchy}

Minimal gravity  $M(2, 2p+1)$ of Lee-Yang series is represented 
in terms of one-matrix model and 
its free energy on a sphere $F_{\rm sphere}$ is given as \cite{BZ}
\be \label{eq1}
	F_{\rm sphere}=\frac{1}{2}\int_0^{u_0} d u\left[\mathcal{P}(u)\right]^2;
	\quad \mathcal{P}(u) =\sum_{m=0} t_m  \frac{u^{m}}{m!}.
\ee
The polynomial equation $\mathcal{P}(u) =0$ is called  the string equation. $u_0$ is one of the solutions of the string equation 
and is related to the free energy  as $u_0 =\pd^2F_{\rm sphere}/{\pd t^2_0}$.

The free energy contains the set of parameters $\{ t_m\}$ 
and becomes the generating function 
because derivatives with respect to  $\{ t_m\}$'s evaluated on-shell provide correlations of the corresponding operators. By on-shell we mean that $t_i = 0$ for all $i$, except $t_{p-1}$ and $t_{p+1}$.  We normalize $t_{p+1}$  to 1, and $t_{p-1}$ is proportional to the cosmological constant $\mu$. Multi-correlation on sphere is given as 
\be
	\ll \prod_i^n O_{a_i} \rr_{{\rm sphere}}
	=  \frac{\pd^n F_{\rm sphere}}{\pd t_{a_1}  \cdots  \pd t_{a_n}}.
\ee
Computing the above quantity on-shell gives us the physical correlation. For example, the two-point correlation 
$\ll O_0 O_0 \rr_{{\rm sphere*}}=u_*$ is the  desired result and further results in
$ \ll O_0 O_{n-1} \rr_{{\rm sphere*}} = {u^{n}_*}/{n!} $, {where the symbol * stands for on-shell value.}
One can also see that the free energy in \eqref{eq1} satisfies the KdV hierarchy on the sphere
\be \label{diff:of:KdVflow}
\frac{\partial^3 F_{\rm sphere}}{\partial t_n \partial t_0^2}=\frac{\pd u}{\pd t_n}=\frac{u^n}{n!}\frac{\pd u}{\pd t_0}.
\ee
The matrix model representation 
is shown to be equivalent to Liouville minimal gravity 
if one takes care of the resonance transformation \cite{BZ}.  

The KdV hierarchy in general is given by
\be\label{Buryak:1.3}
\frac{1}{\lambda^{2}}\frac{2n+1}{2}\frac{\pd^3 F^c}{\pd t^2_0\pd t_n}
=\frac{\pd^2 F^c}{\pd t^2_0}\frac{\pd^3 F^c}{\pd t^2_0\pd t_{n-1}}
+\frac{1}{2}\frac{\pd^3 F^c}{\pd t^3_0}\frac{\pd^2 F^c}{\pd t_0\pd t_{n-1}}
+\frac{1}{8}\frac{\pd^5 F^c}{\pd t^4_0\pd t_{n-1}} \, . 
\ee
where $n\geq 1$. The string equation is
\be\label{Buryak:1.2}
\frac{\pd F^c}{\pd t_0}=\sum_{n\geq0}t_{n+1}\frac{\pd F^c}{\pd t_n}+\frac{t^2_0}{2\lambda^2}.
\ee
The free energy has the genus expansion
\be\label{gen:exp:c}
F^c=\sum_{g=0}^\infty \lambda^{2g-2}F^c_{(g)},
\ee
where {$F^c_{(0)}=F_{\rm sphere}$} and $\lambda$ is a formal expansion parameter. The sphere KdV and string equations can be deduced by considering the dominant part of the general equations. (The equation  $\mathcal{P}(u) =0$ is obtained by a combination of \eqref{Buryak:1.2} and \eqref{gen:exp:c} for $g=0$.)

Witten's conjecture for intersection theory was proved by Kontsevich \cite{Konts} using the one-matrix model.
The generating function for minimal gravity on $g=0,1,2$ has also been constructed using the KdV hierarchy \cite{BT, BBT}, and the resulting correlations (but off-shell, i.e., with arbitrary $t_k$ parameters) have been shown to obey the recursion relations of topological gravity, as suggested by Witten. (In the string equation, $t_1$ is to be shifted by 1 for the comparison of the two cases.)  

\section{Minimal gravity on a disk and KdV hierarchy}

A similar KdV hierarchy (``open KdV hierarchy") has been proposed for intersection theory on the moduli space of Riemann surfaces  with boundary, using an additional flow parameter $s$.
The flow along $t_n$ is given as  \cite{PST}
\be\label{Buryak:1.9}
	\frac{2n+1}{2}\frac{\pd F^o}{\pd t_n}
	=\lambda\frac{\pd F^o}{\pd s}\frac{\pd F^o}{\pd t_{n-1}}
	+\lambda\frac{\pd^2 F^o}{\pd s\pd t_{n-1}}
	+\frac{\lambda^2}{2}\frac{\pd F^o}{\pd t_0}\frac{\pd^2 F^c}{\pd t_0\pd t_{n-1}}
	-\frac{\lambda^2}{4}\frac{\pd^3 F^c}{\pd t^2_0\pd t_{n-1}},
	\qquad n\geq1.
\ee
The  open string equation  is given by
\be\label{Buryak:1.7}
	\frac{\pd F^o}{\pd t_0}=\sum_{n\geq0}t_{n+1}\frac{\pd F^o}{\pd t_n}+\frac{s}{\lambda}.
\ee
 The open KdV together with the string equation is shown to be equivalent to the Virasoro constraints \cite{Buryak}. 

In this section, we explicitly check that the open KdV equations are satisfied by the free energy on a disk. The free energy with a $(1,1)$ boundary is given by \cite{MSS, IR}
\be\label{Gen.Funct.Disk}
	F_{\rm disk}=\sqrt{\pi}\int_0^\infty \frac{dl}{l^{3/2}}e^{-l\mu_B}\int_{t_{0}}^\infty dx \ e^{-lu},
\ee
where $\mu_B$ is the boundary cosmological constant. 
In \eqref{Gen.Funct.Disk}, $u$ satisfies the string equation and KdV on a sphere and  is, 
therefore,  {$u=u(x,t_{n>0})$} a function of {$x$ and $t_n$'s $(n>0)$ but independent of $t_0$.}

The  genus expansion of the free energy with a boundary
\be\label{gen:exp:o}
	F^o=\sum_{g=0}^\infty \lambda^{g-1}F^o_{(g)},
\ee
shows that $F^o_{(0)}$  
satisfies the open KdV at this order:
\be
	\label{def:s-flow}
	\frac {2n+1}2  \frac{\pd F^o_{(0)}}{\pd t_n}
	= 
	\frac{\pd  F^o_{(0)}}{\pd s}\frac{\pd F^o_{(0)}}{\pd t_{n-1}}+
	\frac{1}{2}\frac{\pd F^o_{(0)}}{\pd t_0}\frac{\pd^2 F^c_{(0)}}{\pd t_0\pd t_{n-1}}.
\ee

The one-point correlation on a disk is non-trivial.
The correlation is found from \eqref{Gen.Funct.Disk} by differentiating with respect to $t_n$, using the KdV on sphere (but staying off-shell):
\be
	\ll O_{n} \rr_{\rm disk}
	=  \frac{\pd F_{\rm disk}}{\pd t_n}
	={-}\frac{\sqrt{\pi}}{n!}\int_0^\infty dl\ e^{-l\mu_B}\int^\infty_{u_0} du \frac{ e^{-lu}  }{\sqrt{l}} u^{n},
\ee
{where $u_0=u(x=t_0,t_{n>0})$ and is the same as the one in \eqref{eq1}.}
Since $u$ has a gravitational dimension, one may set $u=u_0\xi$ so that $u_0$ is dimensionful while $\xi$ is dimensionless. 
The integration over $u$ results in the incomplete gamma function.
However, it is more convenient to rewrite the monomial $\xi^n$
as a linear combination of  Legendre polynomials $P_k$:
\be
	\xi^n=\sum_{k=n,n-2,\dots\geq0} (2k+1)\, n!  \, a_{n,k}\,  P_k(\xi)
\ee
where \be
	a_{n,k}=\frac{1}{2^{(n-k)/2}((n-k)/2)!(n+k+1)!!}.
\ee
Then the integration over $\xi$ is given as the  modified Bessel function $K_{n}$ of the second kind:
\be\label{bessel_over_l:integ}
	  \int_1^\infty d\xi  {e^{-u_0 l\xi}} P_k(\xi)=\sqrt{\frac{2}{\pi u_0 l }} {K_{1/2+k}( u_0 l)} .
\ee 
Further, its integration over $l$ is performed (after analytic continuation if necessary) to give 
\be
	\int_0^\infty \frac{dl}{l}\ e^{-l\mu_B} K_{1/2+k}(u_0 l)=\frac{2\pi(-1)^{k+1}}{2k+1}\cosh((1/2+k)\tau),
\ee
where we put\footnote
{The boundary parameter $\mu_B$ is independent of  KdV parameters 
but $\tau$ depends on KdV parameters due to $u_0$.}
$\mu_B/u_0=\cosh(\tau)$.
Therefore, the correlation number is given as follows in terms of the Chebyshev polynomial $T_n (\cosh(x)) = \cosh(nx) $:
\be\label{disk:corr:numb}
	\ll O_{n} \rr_{\rm disk}
	=2\pi\sqrt{2}u_0^{n+1/2}(-1)^{n+1}\sum_{k=n,n-2,\dots\geq0}a_{n,k}T_{2k+1}(\cosh(\tau/2)).
\ee 
If one uses the identity of Chebyshev polynomials $T_{n+2}=2T_2T_n-T_{|n-2|}$
and the two-point correlation on sphere $ 	\ll O_0 O_{n-1} \rr_{\rm sphere}  	=  {u^{n}_0}/{n!}$, 
 one has  the following recursive relation 
\be\label{disk:corr:numb:rec}
	\frac{2n+1}{2}\ll O_{n} \rr_{\rm disk}
	=-u_0 T_2(\cosh(\tau/2))\ll O_{n-1} \rr_{\rm disk}+\frac{1}{2}\ll O_{0} \rr_{\rm disk}\ll O_0 O_{n-1} \rr_{\rm sphere}.
\ee
Comparing the recursion  in \eqref{disk:corr:numb:rec} 
with the open KdV on disk in  \eqref{def:s-flow}, 
one concludes that the free energy on a disk follows the open KdV and 
the flow along $s$ reads:
\be \label{dFds}
	{\frac{\pd F^o_{(0)}}{\pd s}}  = -u_0 T_2(\cosh(\tau/2)) 	=u_0 \cosh(\tau) = -\mu_B,
\ee
{Considering \eqref{def:s-flow}, \eqref{disk:corr:numb:rec} and \eqref{dFds} we arrive at the conclusion that $F^o_{(0)}$ and $F_{\rm disk}$ are related by the Legendre transformation  
\be
	F^o_{(0)}(s)=F_{\rm disk}(\mu_B) -\mu_B\,s.
\ee}
We note that the above calculations are done off-shell and 
therefore, \eqref{dFds} holds off-shell also. Using this result, one can prove that 
the open KdV on the disk holds for $F_{\rm disk}$ (and multi-correlations) 
 using just
the integral representation of $F_{\rm{disk}}$ in \eqref{Gen.Funct.Disk} 
and the fact 
$-\mu_Be^{-l\mu_B} = \tfrac{\partial e^{-l\mu_B}}{\partial l}$. It is worth pointing out that the result of \cite{PST, Buryak} for $g=0$
\be
{\frac{\partial F^o_{(0)}}{\partial s} 
= \frac{1}{2}\left(\frac{\partial F^o_{(0)}}{\partial t_0}\right)^2 + \frac{\partial^2 F^c_{(0)}}{\partial t_0^2}}
\ee
still holds,\footnote{
{To make it match exactly, we may rescale $F_{\rm{disk}}\to cF_{\rm{disk}}$ with $c^2=-1/(2\pi^2)$.}}
essentially because of the Chebyshev polynomial identity $T_2 = 2T_1^2 - 1$.
\section{Conclusion}
We demonstrate that the open KdV hierarchy holds for the matrix models of the Lee-Yang series of the minimal gravity on disk.
In order to prove this, the fact that the $s$ flow of the generating function on a disk is governed by the boundary cosmological constant $\mu_B$ is essential. This result can be compared with \cite{Ale2,Ale}, where intersection theory with boundary is investigated using a Penner-type matrix model, but has no boundary parameter. One could investigate more general boundary conditions and boundary correlations on the disk, and the corresponding KdV hierarchies. It might be interesting to investigate if the $s_n$ flows of \cite{Buryak,Ale2, Buryak2} have an interpretation in the Lee-Yang series, {which would be related to the other boundary condition than the simple (1,1) boundary condition considered in the text.}

The minimal gravity $M(2, 2p+1)$ is described by one variable $u$ which is 
the coordinate of $A_1$ Frobenius manifold \cite{BDM}. 
It is known that its dual $A_{2p}$ Frobenius manifold (which has $2p$ number of coordinates $\{ u^a\}$)
can describe the same correlation on sphere. 
However, the correlation on disk  is not fully understood in terms of the dual manifold
due to the non-canonical nature of the  integral representation of the free energy \cite{ABR, BMR}. 
The open KdV hierarchy can be a guide to define the free energy on disk 
whose details will be considered in a separate paper.

\subsection*{Acknowledgements}
The work was partially supported by National Research Foundation of Korea grant number 2017R1A2A2A05001164.



\begin{thebibliography}{99}
\bibitem{Witten}
E. Witten, 
\emph{Two dimensional gravity and intersection theory on moduli space}, 
{Surveys in Diff. Geom. 1} (1991) 243-310.

\bibitem{DVV}
R. Dijkgraaf, H. Verlinde and E. Verlinde,
\emph{Loop equations and Virasoro constraints in nonperturbative 2-D quantum gravity}, 
{Nucl. Phys.} {B348} (1991) 435-456.

\bibitem{MMM}
A. Marshakov, A. Mironov and A. Morozov,
\emph{On equivalence of topological and quantum 2-d gravity},
{Phys. Lett.} {B274} (1992) 280-288, {[arXiv:hep-th/9201011]}.

\bibitem{Kostov}
V. A. Kazakov, I. K. Kostov and A. A. Migdal ,
\emph{Critical Properties of Randomly Triangulated Planar Random Surfaces},
Phys. Lett. B157 (1985) 295-300.

\bibitem{Kazakov}
V. A. Kazakov,
\emph{Ising model on a dynamical planar random lattice: Exact solution},
Phys. Lett. A119 (1986) 140-144.

\bibitem{KPZ}
V. G. Knizhnik, A. M. Polyakov and A.B. Zamolodchikov,
\emph{Fractal Structure of 2D Quantum Gravity},
Mod. Phys. Lett. A3 (1988) 819.

\bibitem{Doug}
M. R. Douglas, 
\emph{Strings less than one dimension and generalized KdV hierarchies},
Phys. Lett. B238 (1990) 176.

\bibitem{Ginsparg}
P. H. Ginsparg, M. Goulian, M. R. Plesser and J. Zinn-Justin,
\emph{(p, q) STRING ACTIONS},
Nucl. Phys. B342 (1990) 539-563.

\bibitem{PST}
R. Pandharipande, J. P. Solomon and R. J. Tessler,
\emph{Intersection Theory on Moduli of Disk, Open KdV and Virasoro}, 
{[arXiv:1409.2191 [math.SG]]}.
 
 \bibitem{Buryak}
A. Buryak,
\emph{Equivalence of the Open KdV and the Open Virasoro Equations for the Moduli Space of Riemann Surfaces with Boundary}, 
{Letters in Mathematical Physics 105 (2015) no. 10 1427-1448}, {[arXiv:1409.3888 [math.AG]]}.

\bibitem{BZ}
A. A. Belavin and A. B. Zamolodchikov,
\emph{On Correlation Numbers in 2D Minimal Gravity and Matrix Models},
J.Phys. A42 (2009) 304004, [arXiv:0811.0450 [hep-th]].

\bibitem{Konts}
M. Konstevich, 
\emph{Intersection theory on the moduli space of curves and the matrix Airy function}, 
{Comm. Math. Phys. 147}  (1992) 1-23.

\bibitem{BT}
A. Belavin and G. Tarnopolsky,
\emph{Two dimensional gravity in genus one in Matrix Models, Topological and Liouville approaches}, 
JETP Lett. 92 (2010) 257-267, {[arXiv:1006.2056 [hep-th]]}.

\bibitem{BBT}
A. Belavin, M. Bershtein and G. Tarnopolsky,
\emph{A remark on the three approaches to 2D quantum gravity}, 
JETP Lett. 93 (2011) 47-51, {[arXiv:1010.2222 [hep-th]]}.

\bibitem{MSS}
G. W. Moore, N. Seiberg and M. Staudacher,
\emph{From loops to states in 2-D quantum gravity}
Nucl.Phys. B362 (1991) 665-709.

\bibitem{IR}
G. Ishiki and C. Rim,
\emph{Boundary correlation numbers in one matrix model},
{Phys. Lett.} {B694} (2011) 272-277, {[arXiv:1006.3906 [hep-th]]}.

\bibitem{Ale2}
A. Alexandrov, 
\emph{Open intersection numbers, matrix models and MKP hierarchy},
JHEP 1503 (2015) 042, {[arXiv:1410.1820 [math-ph]]}.

\bibitem{Ale}
A. Alexandrov, 
\emph{Open intersection numbers, Kontsevich-Penner model and cut-and-join operators},
JHEP 1508 (2015) 028, {[arXiv:1412.3772 [hep-th]]}.

 \bibitem{Buryak2}
A. Buryak,
\emph{Open intersection numbers and the wave function of the KdV hierarchy}, 
{Moscow Math. J. 16 (2016) no.1, 27-44}, {[arXiv:1409.7957 [math-ph]]}.

\bibitem{BDM}
A. Belavin, B. Dubrovin and B. Mukhametzhanov,  
\emph{Minimal Liouville Gravity correlation numbers from Douglas string equation}, 
{JHEP} {01} (2014) 156, {[arXiv:1310.5659 [hep-th]]}.

\bibitem{ABR}
K. Aleshkin, V.  Belavin and C. Rim,
 \emph{	Minimal gravity and Frobenius manifolds: bulk correlation on sphere and disk }, 
{JHEP} { 1711}  (2017) 169, {[arXiv:1708.06380 [hep-th]]}.

\bibitem{BMR}
A. Bawane, H. Muraki and C. Rim,  
\emph{Dual Frobenius manifolds of minimal gravity on disk}, 
{JHEP} {1803} (2018) 134, {[arXiv:1801.10328 [hep-th]]}.

\end{thebibliography}
\end{document}